\documentclass[twocolumn]{aastex631}
\usepackage{amsmath}
\newcommand{\eb}{\begin{equation}}
\newcommand{\ee}{\end{equation}}
\newcommand{\masyr}{mas yr$^{-1}$}
\newcommand{\uasyr}{$\mu$as yr$^{-1}$}

\definecolor{rkka}{RGB}{219,66,32}

\shorttitle{Systemic proper motions of quasars}
\shortauthors{Makarov \& Secrest}

\begin{document}

\title{Testing the Cosmological Principle: Astrometric Limits on Systemic Motion of Quasars at Different Cosmological Epochs}

\correspondingauthor{Valeri V. Makarov}
\email{valeri.makarov@gmail.com}

\author[0000-0003-2336-7887]{Valeri V. Makarov}
\affiliation{U.S. Naval Observatory, 3450 Massachusetts Ave NW, Washington, DC 20392-5420, USA}

\author[0000-0002-4902-8077]{Nathan J. Secrest}
\affiliation{U.S. Naval Observatory, 3450 Massachusetts Ave NW, Washington, DC 20392-5420, USA}

\begin{abstract}
A sample of $60,410$ bona fide optical quasars with astrometric proper motions in Gaia EDR3 and spectroscopic redshifts above 0.5 in an oval 8400 square degree area of the sky is constructed. Using orthogonal Zernike functions of polar coordinates, the proper motion fields are fitted in a weighted least-squares adjustment of the entire sample and of six equal bins of sorted redshifts. The overall fit with 37 Zernike functions reveals a statistically significant pattern, which is likely to be of instrumental origin. The main feature of this pattern is a chain of peaks and dips mostly in the R.A.\ component with an amplitude of 25~\uasyr. This field is subtracted from each of the six analogous fits for quasars grouped by redshifts covering the range 0.5 through 7.03, with median values 0.72, 1.00, 1.25, 1.52, 1.83, 2.34. The resulting residual patterns are noisier, with formal uncertainties up to 8~\uasyr\ in the central part of the area. We detect a single high-confidence Zernike term for R.A.\ proper motion components of quasars with redshifts around 1.52 representing a general gradient of 30 \uasyr\ over $150\degr$ on the sky. We do not find any small- or medium-scale systemic variations of the residual proper motion field as functions of redshift above the $2.5\,\sigma$ significance level. 
\end{abstract}

\section{Introduction} \label{intr.sec}
It is generally assumed that the Hubble flow of the expanding Universe has a time-dependent radial component but no time-dependent tangential component. The existence of such a non-radial component would imply that observers located at different parts of the universe at earlier cosmological epochs would see an asymmetric distribution of radial expansion on the sky. This can be taken as an observable violation of the cosmological principle, extended to the time domain, which asserts that the observed motions of distant quasars do not exhibit a position dependence beyond that induced by the local effects of relativistic aberration and Doppler boosting. Using statistical and astrometric terms, the testable null hypothesis based on this principle is that a sufficiently large sample of quasars in a sufficiently large area of the sky is expected to have zero systemic tangential motion. The full-sky distribution quasars exhibits a dipole component significantly larger than predicted given a purely kinematic interpretation of the cosmic microwave background dipole, apparently violating the cosmological principle \citep{2021ApJ...908L..51S}. In this Letter, we develop a unique test of the cosmological principle that uses precision astrometry of a large number of spectroscopically confirmed quasars.

The advancement of space astrometry, culminating in the European Space Agency's Gaia mission \citep{2016A&A...595A...1G}, opens up new pathways to testing this basic assumption. The intersection of the Gaia Early Data Release 3 \citep[EDR3 hereafter, ][]{2021A&A...649A...1G} with the mid-infrared quasars and AGNs detected by WISE \citep{2015ApJS..221...12S} counts over $600,000$ sources with measurements of proper motions and parallaxes. These astrometric measurements are expected, in general, to deviate from zero only because of observational and statistical errors. This Bayesian prior has been used to construct the fundamental optical reference frame Gaia-CRF. Indeed, the indeterminate rigid rotation of the proper motion system has been removed by constraining the overall spin of a large all-sky ensemble of quasars to zero \citep{2018A&A...616A..14G}. This helps to make the Gaia-CRF as closely inertial as possible. The desired quality of the reference system comes at the cost of being in part predicated on the cosmological principle and the standard cosmological model.

The main purpose of this study is to test if large-scale systemic patterns of tangential motion can be found at the early stages of cosmic expansion with available astrometric and spectroscopic data. Our approach is to collect a large sample of optical quasars with accurate spectroscopic redshifts from the Sloan Digital Sky Survey (SDSS) and proper motion measurements from Gaia Early Data Release 3 (EDR3). The zero-points of astrometric proper motions are corrected for small biases and their formal errors are inflated by a factor of 1.06 to make the sample distribution consistent with the theoretically expected probability distribution (Makarov \& Secrest 2022, in prep.). The observed proper motion field of sources within a large elliptical area of the sky is fitted with a limited set of 2D Zernike functions. The resulting pattern is considered to represent systematic (sky-dependent) astrometric errors of the Gaia mission. The fitting procedure is repeated for six subsamples of quasars corresponding to six non-overlapping intervals of redshift. The instrumental pattern for the general sample is subtracted from each fit and the residual maps are investigated for statistically significant signals.

\section{Astrometric Quasar Sample}

We start with a collection of $621,946$ sources from the MIR-selected quasars from \citep{2015ApJS..221...12S} cross-matched with Gaia EDR3 within a circle with an angular radius of $0\farcs5$. We downselect only objects with spectroscopically measured redshifts from the SDSS-IV \citep{2017AJ....154...28B} \texttt{specObj-dr16.fits} table\footnote{\url{https://www.sdss.org/dr16/spectro/spectro_access/}} of high quality, as detailed in Makarov \& Secrest (2022). We also remove potential stellar interlopers by applying the filter \texttt{parallax}/\texttt{parallax\_error}~$<4$. These two cuts leave $125,933$ sources that are bona fide quasars with valid proper motion measurements.\footnote{A valid measurement means that the measurement has an error estimate, not that the measured proper motion is significant.} The next cut is at redshift $z>0.5$, leaving $107,568$ objects. This filter removes AGNs hosted by relatively nearby galaxies that are resolved by Gaia, which introduces unmodeled variance in the astrometric solution (Makarov \& Secrest 2022, in prep.). We further apply a filter at the Gaia quality parameter \texttt{phot\_bp\_rp\_excess\_factor} $<1.4$, which may help remove double AGNs, optical pairs, and multiply imaged lenses with perturbed astrometry. The proper motions are corrected with zero-point offsets of  $-3$~and $+2$ \masyr\ and the standard errors are inflated by factors 1.056 and 1.064 for R.A.\ and decl.\ components, respectively. These corrections make the overall sample of normalized proper motions close to ${\cal N}(0,1)$. Our empirically determined error correction factors are consistent with the standard deviation of normalized proper motions of 1.063 determined by \citet{2021A&A...649A...9G}. 

Even after these cuts, there is a small admixture of sources with elevated proper motions. This can be seen, for example, from the highest quantiles of the absolute proper motion in RA: $\{2.252, 2.735, 8.015\}$
\masyr\ at $\{0.998,0.999,1.0\}$. Because of the subsequent weighting with the (inflated) formal errors, we want to get rid of statistical outliers and a few remaining stars, which can affect the estimated systemic patterns. This is achieved by removing a few percent of the working sample with normalized proper motion components in excess of 1.96. 

\begin{figure}
    \includegraphics[width=\columnwidth]{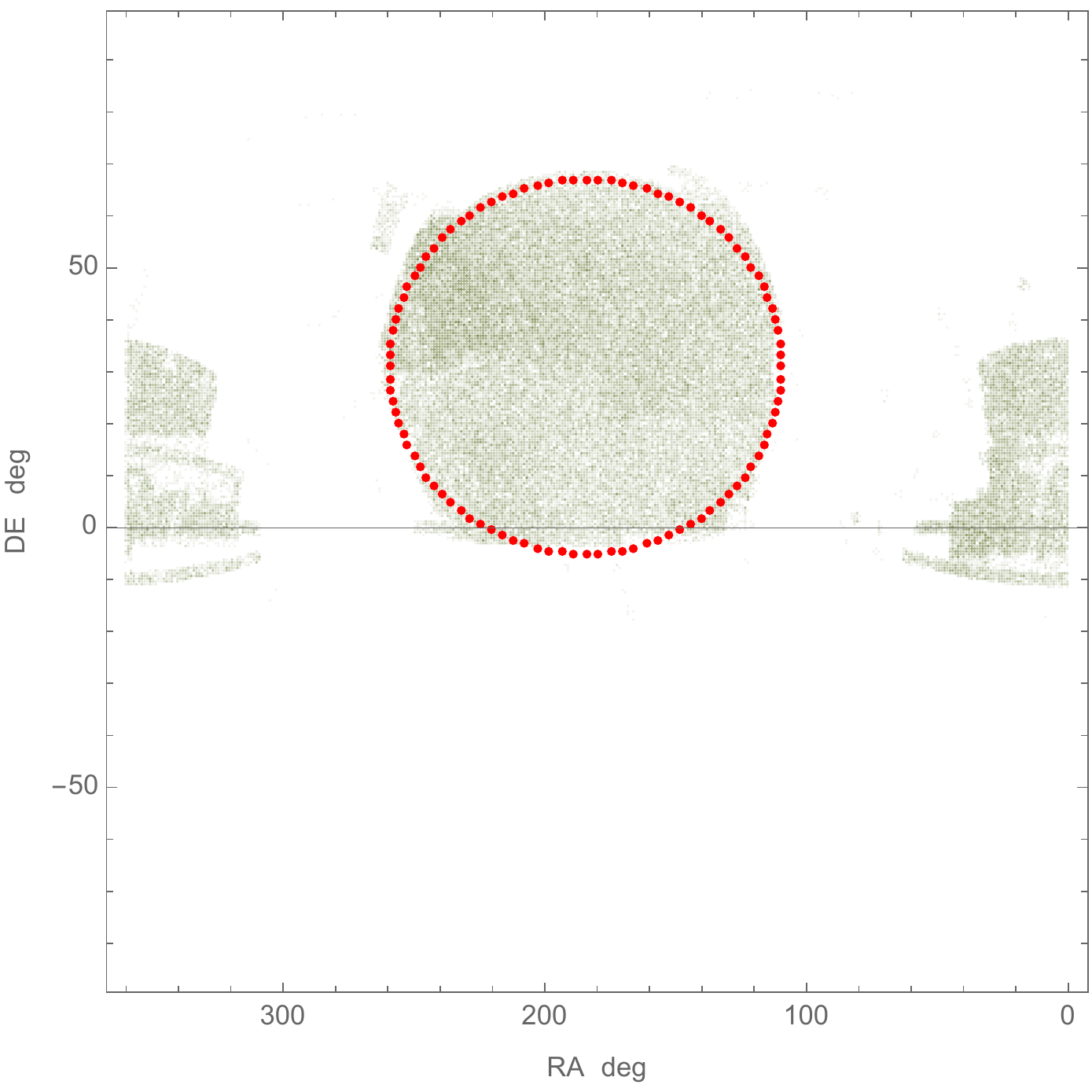}
    \caption{Sky footprint of spectroscopic quasars from SDSS with valid proper motion measurements in Gaia EDR3. The red dotted line indicates the boundary of the elliptical area selected for this study.}
    \label{foot.fig}
\end{figure}

The distribution of resulting quasar sample on the sky is shown in Fig. \ref{foot.fig}. Since we are after large scale sky-correlated signals, only the largest contiguous area can be used. The dotted red line indicates the boundary of a contiguous area where most of the SDSS sources are located, selected for this study. The center of this area is at $(\alpha_0,\delta_0)=(184\fdg4,31\fdg0)$, with semi-major axes $(s_\alpha , s_\delta)=(75\degr,36\degr)$. Note that the selected area is roughly elliptical on the celestial sphere and is extended in the Right Ascension direction. We compute normalized angular coordinates and polar coordinates for each quasar as
\begin{eqnarray}
x &= & (\alpha - \alpha_0)/ s_\alpha \nonumber\\
y &= & (\delta - \delta_0)/ s_\delta \nonumber\\
r &= & \sqrt{x^2+y^2}\nonumber\\
\phi &= & \arctan (x,y).
\end{eqnarray}
Only quasars with $r<1$ are accepted, yielding a final working sample of $60,410$ objects. The covered area is roughly 8400 square degrees. 

\section{Computations}

Zernike functions $Z_n^m(r,\phi)$, sometimes called odd and even Zernike polynomials, are orthogonal scalar functions representing a basis of functions defined on a unit disk.\footnote{\url{https://mathworld.wolfram.com/ZernikePolynomial.html}} They are composed of the radial Zernike polynomials of degree $n$ and Fourier harmonics of order $m$, $n\ge m\ge 0$. These functions are often used in optical engineering to describe optical aberrations \citep{born1959principles} and in astrometry to represent instrumental distortions in circular fields of view \citep{2012PASP..124..268M}. 
We use 37 lower-order Zernike functions ordered according to the Noll's scheme, which is the complete set up to $n=7$. Going for higher degrees and wave numbers is not justified lest our results become affected by the small-scale structure at the level of superclusters. There is no specific correspondence between the Zernike function degree and a characteristic wavenumber, but we note that up to 7 waves in azimuth $\phi$ are present at $n=7$. Very roughly, the characteristic wavelength captured by this decomposition is $55\degr$.

The functional fit is computed by a least-squares solution for 37 coefficients separately for R.A.\ and decl.\ proper motion components. Each of the $60,410$ equations is weighted by the corrected standard error. We note that a full least-squares adjustment is required, because the orthogonality of the continuous Zernike functions is not preserved on the discrete set of sources due to the non-uniform distribution of the number density and weights. The former can be seen by eye in Fig. \ref{foot.fig}. The products of the two solutions are two sets of coefficients in units of \masyr\ and their $37\times 37$ covariance matrices. The square root of the diagonal provides the formal uncertainties of the fit.

\section{Results}
\subsection{Full Sample}
\begin{figure*}
    \includegraphics[width=0.5\textwidth]{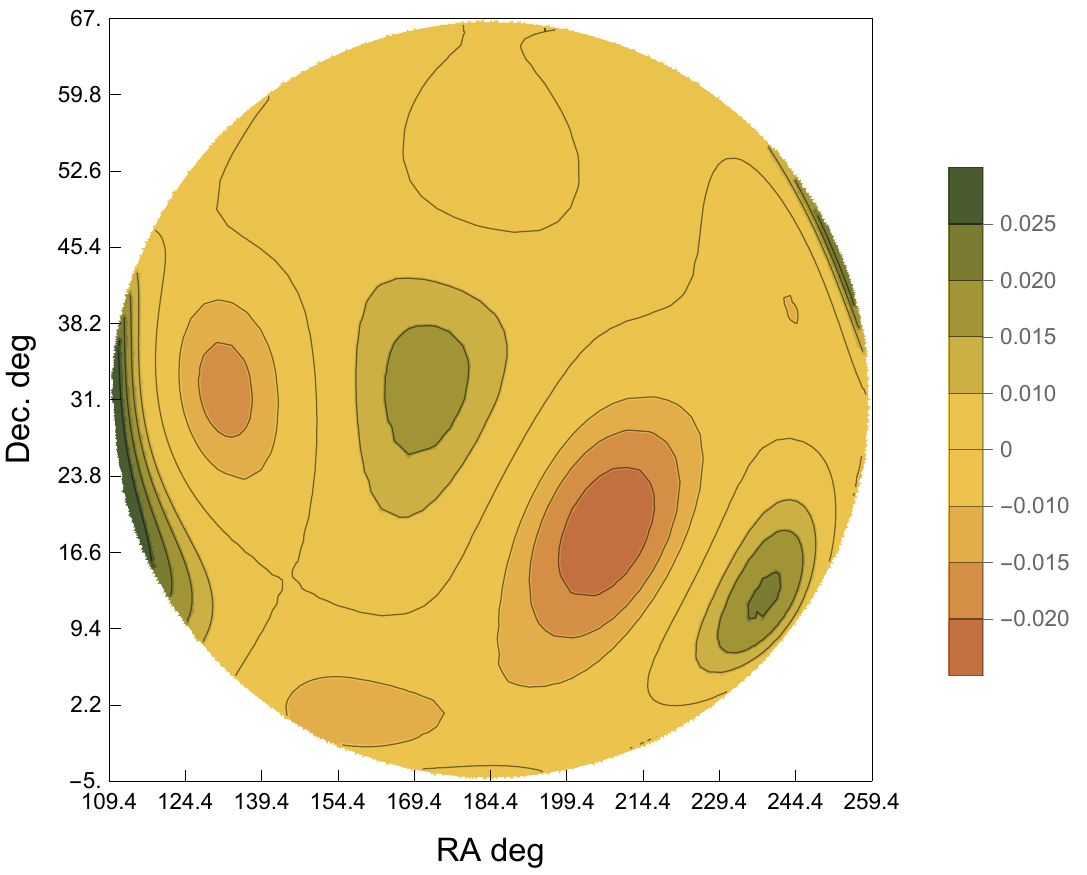}
    \includegraphics[width=0.5\textwidth]{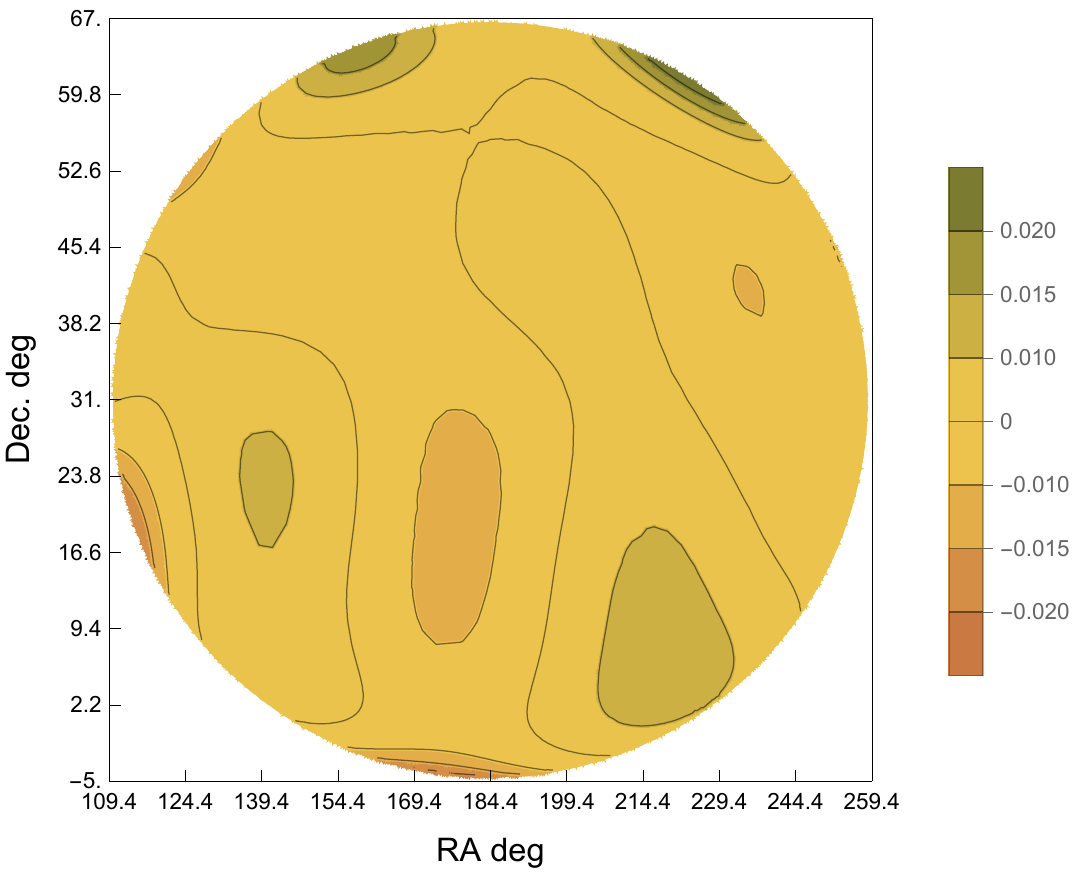}
    \caption{Contour maps of proper motion fields in R.A.\ (left) and decl.\ (right) coordinates fitted with 37 Zernike functions each. The values are in \masyr. The coordinates are in degrees. Note the direction of the R.A.\ axis increasing from left to right.}
    \label{all.fig}
\end{figure*}

The Zernike function fit of the general sample of $60,410$ quasars in the selected area of the sky essentially represents the smoothed proper motion field. The value of this field for any point within the area is computed from the known values of the fitting functions and their estimated coefficients. The result is displayed in Fig. \ref{all.fig} as a color-coded map with contours of equal levels. The amplitude of proper motion variation in the R.A.\ component (left plot) is slightly larger than that in the decl.\ component (right plot) reaching 25 \uasyr in the central part. The main feature is a chain of dips and peaks stretching mostly in the east-west direction. A trace of a similar feature seems to be present for the decl.\ component but it is less pronounced. Is this pattern statistically significant? The standard error of the fit for each point $\{r,\phi\}$ is computed as
\eb 
\sigma^{(Z)}(r,\phi)=\left[\boldsymbol{Z}^T(r,\phi)\;\boldsymbol{C}^{(Z)}\;\boldsymbol{Z}(r,\phi)\right]^\frac{1}{2},
\ee 
where $\boldsymbol{C}^{(Z)}$ is the covariance matrix of the solution vector and $\boldsymbol{Z}(r,\phi)$ is the vector of Zernike function values.

\begin{figure*}
    \includegraphics[width=0.5\textwidth]{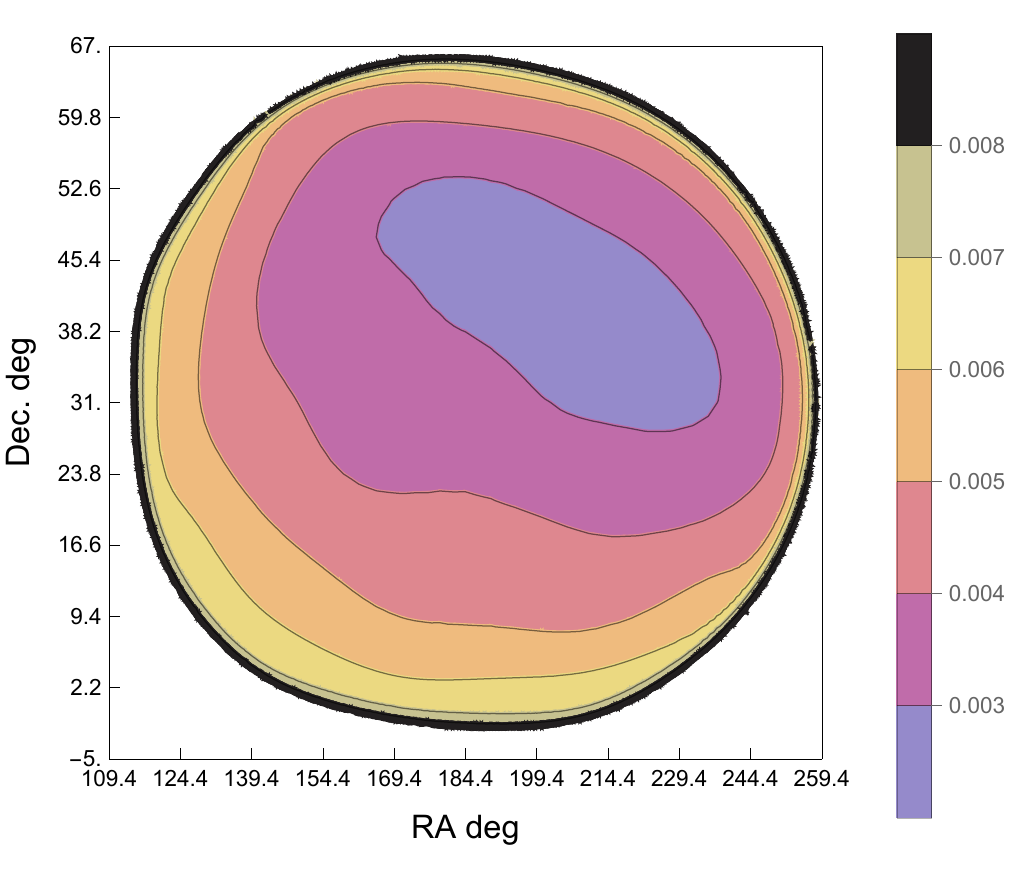}
    \includegraphics[width=0.5\textwidth]{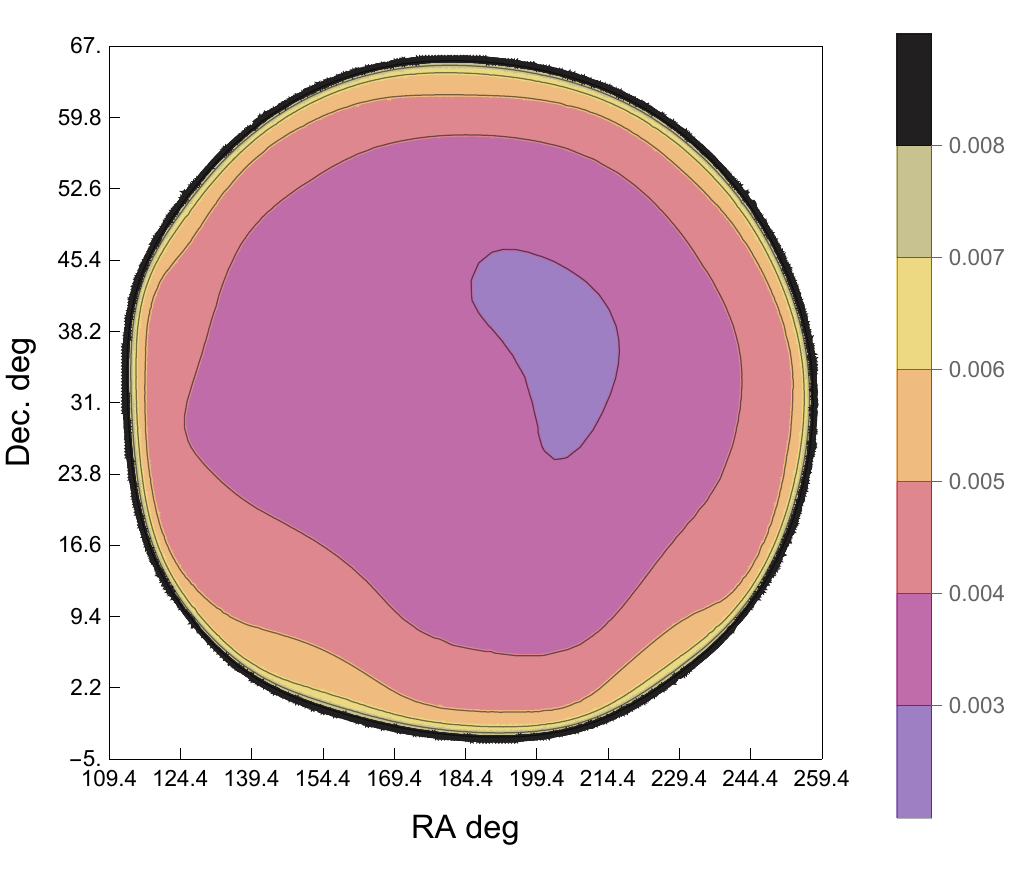}
    \caption{Contour maps of standard errors of the component proper motion fields in R.A.\ (left) and decl.\ (right) shown in Figure~\ref{all.fig}. The values are in \masyr. The coordinates are in degrees. Note the direction of the R.A.\ axis increasing from left to right.}
    \label{err.fig}
\end{figure*}

The resulting distribution of the error of the fit is shown in Fig. \ref{err.fig}. The best precision of 3 \uasyr\ or better is achieved in small patches near the center where the density of quasars is the highest. The fit performance deteriorates toward the edges of the sampled area, which is a feature of the Zernike radial polynomials. The sharp variations near the boundary seen in Fig. \ref{all.fig} are therefore not of high significance. The overall reliability of the fit can also be quantified by the robustness of a least-squares solution introduced in \citet{2021AJ....161..289M}, which depends on the spectrum of singular values of the weighted design matrix. This value comes up to ${\cal R}=21.9$, which is a high robustness indicating a factor of $\sim 1/2$ reduction with respect to the absolute maximum value ${\cal R}_{\rm max}=40.4$ for this problem.

An alternative way of estimating the significance of the fit is to consider the S/N values of the Zernike function coefficients, 
\eb 
\chi^{(Z)}_i = a^{(Z)}_i /\sqrt{C_{ii}},
\ee 
where $a^{(Z)}_i$, $i=1,2,\ldots,37$, are the estimated coefficients. These values are expected to be drawn from ${\cal N}(0,1)$ in the null hypothesis (no signal). We find three terms for the R.A.\ fit with $|\chi|$ values in excess of 3, corresponding to a confidence of 0.99865, namely, $Z_{11}=\sqrt{5} (1 - 6 \,r^2 + 6 \,r^4)$, $Z_{16}= 
 2 \sqrt{3} (3 \,r - 12 \,r^3 + 10 \,r^5) \,\cos{\phi}$, and $Z_{23}= 
 \sqrt{14} (6 \,r^2 - 20 \,r^4 + 15 \,r^6) \sin{2 \phi}$. Together, they are responsible for the pattern seen ion Figure~\ref{all.fig}, left. The declination fit, on the other hand, does not have any terms with $|\chi|$ values exceeding 2.45. The $p$-values from statistical testing of the sample distribution are, for the R.A.\ fit, 0.0014 from the Anderson-Darling test and 0.00018 from the Pearson $\chi^2$ test, whereas they are 0.094 and 0.14, respectively, for the decl.\ component.  The largest S/N in terms of absolute value is $-6.2$ and is achieved for the most prominent blob at $\{207\degr,+18\degr\}$.

\subsection{Residual proper motion fields at different redshifts}

The same least-squares fitting procedure is repeated for six subsets of the general sample grouped by redshift. The intervals of $z$ and the median values are specified in Table \ref{bins.tab}. Each subset includes 1/6 of the sample, i.e., about $10,068$ quasars. The robustness of these fits varies between 8.3 and 9.0. The numbers are smaller than the general robustness because of the smaller number of condition equations, while the reduction with respect to the theoretical maximum of 16.2 is still about $1/2$. This proves that no additional structural weakness is introduced in the solution by binning the sample into 6 subsamples.

\begin{deluxetable}{lccr} \label{bins.tab}
\tablehead{\colhead{bin} & \colhead{Min[$z$]} & \colhead{Max[$z$]} & \colhead{Median[$z$]} }
\startdata
1 & 0.50 & 0.87 & 0.72 \\
2 & 0.87 & 1.13 & 1.00 \\
3 & 1.13 & 1.38 & 1.25 \\
4 & 1.38 & 1.66 & 1.52 \\
5 & 1.66 & 2.05 & 1.83 \\
6 & 2.05 & 7.03 & 2.34  
\enddata
\caption{Redshift boundaries and median values for the 6 subsets of quasars.}
\end{deluxetable}

Once the proper motion field is fitted for each redshift subset, the general sample fit is subtracted yielding a residual fit. This subtraction is performed directly in the space of Zernike coefficient vectors. The covariance of the residual vector is assumed to be the sum of the covariances for the general sample and the given bin. The small positive correlation between the two solutions is ignored, resulting in a systematic overestimation of uncertainties. Using these data, S/N values can be computed for the fit coefficients and for the residual proper motion fields. Generally, the formal uncertainties of Zernike coefficients increase by a factor of $\sqrt{6}$ with respect to the overall sample, from 0.9 \uasyr\ to 2.2 \uasyr.
 
Across the six redshift subsets, we found only one Zernike function term (out of 444 fitted) with a S/N value above 3. This term is $Z_2=2\,r\,\cos{\phi}$ with a $\chi=-3.226$ for the residual R.A.\ component of bin 4, which has a median redshift of 1.52. The corresponding pattern is a gradient of proper motion in R.A.\ across the sampled area from $-15$ \uasyr\ at the east end to $+15$ \uasyr\ at the west end. The formal confidence of this trend is 0.99937.  The next most significant terms are $Z_1=1$ with a $\chi=2.52$ for decl.\ proper motions in bin 1 (i.e., a constant positive offset of 5 \uasyr) and $Z_{18}=2\, \sqrt{3} (-4 \,r^3 + 5\, r^5)\, \cos{3 \,\phi}$ with a $\chi=-2.37$ for decl.\ proper motions in bin 2.

 \section{Discussion}
 
We collected a sample of $60,410$ distant quasars with accurate spectroscopic redshifts from the SDSS and proper motions from Gaia in a contiguous oval area of approximately $8400$ square degrees (1/5 of the sphere). Fitting 37 Zernike functions of scaled polar coordinates to this general sample resulted in a statistically significant pattern of proper motions characterized by a string of alternating dips and peaks mostly in the RA component separated by approximately $38\degr$ on the sky. The most prominent features reach S/N values of 6 and higher and correspond to proper motion deviations up to 25 \uasyr. This pattern is almost certainly of instrumental origin and represents a sky-correlated systematic error in the Gaia data. We subtract this artefact from each of six analogous fits for proper motions of quasars divided into non-overlapping subset by redshifts covering the range from 0.5 to 7.03. The residual proper motion fields, unfortunately, have greater formal uncertainties up to $\sim 8$ \uasyr\ in the center of the field. 

For the redshift-selected subsets, we find only one Zernike term that can be considered highly significant. This term represents a linear gradient of the R.A.\ component for quasars around $z=1.52$ (bin 4) in the R.A.\ direction from $-15$ to $+15$ \uasyr\ across the sampled area. Taking into account the full extent of $150\degr$, this pattern can only represent a large-scale anomaly, if it is real. The other detected signals are of much lower confidence. The achieved sensitivity can be quantified by the median length of the fitted proper motion vector field, which varies between 12 and 16 \uasyr\ for the six samples of redshift. With the cosmological parameters of concordance $\Lambda$CDM \citep{2020A&A...641A...6P}, the corresponding characteristic tangential velocity ranges between $1.0\times10^5$ and $1.3\times10^5$ km~s$^{-1}$ at the median redshift values (Table \ref{bins.tab}). The upper limit on any transverse motion exhibited by the quasars in our sample is therefore $\lesssim10^5$~km~s$^{-1}$.

As galaxy groups can exhibit bulk motions of up to $\lesssim1000$~km~s$^{-1}$ without being significantly in tension with the cosmological principle \citep[e.g.,][]{2011MNRAS.414..264C, 2020arXiv200310420M}, our results show that a future astrometric mission will need to accomplish about a factor of $\sim100$ improvement in astrometric precision to be useful for such tests of concordance cosmology. This may be possible if combining data from a future Gaia-like mission such as GaiaNIR with current Gaia data \citep[e.g.,][]{2021ExA....51..783H}. We have therefore presented in this work a mathematical formalism and methodology for performing an important test of the cosmological principle once such data become available \citep[around 2045;][]{2021arXiv210707177H}.

While this astrometric test of the cosmological principle will have to wait a couple of decades, we note that even the large limits we have placed on bulk transverse motion in our sample would otherwise have to be assumed. In a spectroscopic sample of galaxies or quasars, the observed cosmological redshift is a function of the true cosmological redshift, the relativistic Doppler effect, and the gravitational redshift. The latter term is of order $10^{-5}$ \citep[e.g.,][]{2011Natur.477..567W}, so the primary degeneracy is with the relativistic Doppler effect. This is expressed as:

\begin{equation} \label{eq:doppler}
1+z = \gamma (1 + \beta \cos\theta)
\end{equation}

\noindent where $\gamma$ is the Lorentz factor, $\beta=v/c$, and $\theta$ is the angle of motion with respect to the line of sight. For transverse motion, $1+z=\gamma$, so spectra appear redshifted due to time dilation. The upper limit our work has set is $\beta\lesssim c/3$, or $\Delta z < 0.06$. This is comparable to the redshift difference, induced by radial motion, at which cluster membership can be discriminated, so even the modest limits set by current data may still be of use for studies of large-scale structure.

High precision astrometry allows for the motion of the universe to be studied directly, opening up entirely new windows into important astrophysical questions, such as the structure and motion of the Galaxy and its satellites. With the advent of Gaia, novel uses of precision astrometry in the extragalactic domain, such as studying the structure of AGN jets \citep[e.g.,][]{2017A&A...598L...1K}, or detecting dual AGNs \citep[e.g.,][]{2021NatAs...5..569S}, have been developed. In this work, we have introduced a new methodology, based on methods developed for optical engineering, to constrain the transverse bulk motions of quasars, which are important ``test particles'' for cosmological studies. It is clear that current and future astrometric capabilities are becoming increasingly important for 21\textsuperscript{st} century astrophysics \citep{2021FrASS...8....9K}.


\begin{acknowledgements}
The authors are grateful to the anonymous referee, whose comments
and corrections helped to improve the paper. 
Funding for the Sloan Digital Sky 
Survey IV has been provided by the 
Alfred P. Sloan Foundation, the U.S. 
Department of Energy Office of 
Science, and the Participating 
Institutions. 

SDSS-IV acknowledges support and 
resources from the Center for High 
Performance Computing  at the 
University of Utah. The SDSS 
website is www.sdss.org.

SDSS-IV is managed by the 
Astrophysical Research Consortium 
for the Participating Institutions 
of the SDSS Collaboration including 
the Brazilian Participation Group, 
the Carnegie Institution for Science, 
Carnegie Mellon University, Center for 
Astrophysics | Harvard \& 
Smithsonian, the Chilean Participation 
Group, the French Participation Group, 
Instituto de Astrof\'isica de 
Canarias, The Johns Hopkins 
University, Kavli Institute for the 
Physics and Mathematics of the 
Universe (IPMU) / University of 
Tokyo, the Korean Participation Group, 
Lawrence Berkeley National Laboratory, 
Leibniz Institut f\"ur Astrophysik 
Potsdam (AIP),  Max-Planck-Institut 
f\"ur Astronomie (MPIA Heidelberg), 
Max-Planck-Institut f\"ur 
Astrophysik (MPA Garching), 
Max-Planck-Institut f\"ur 
Extraterrestrische Physik (MPE), 
National Astronomical Observatories of 
China, New Mexico State University, 
New York University, University of 
Notre Dame, Observat\'ario 
Nacional / MCTI, The Ohio State 
University, Pennsylvania State 
University, Shanghai 
Astronomical Observatory, United 
Kingdom Participation Group, 
Universidad Nacional Aut\'onoma 
de M\'exico, University of Arizona, 
University of Colorado Boulder, 
University of Oxford, University of 
Portsmouth, University of Utah, 
University of Virginia, University 
of Washington, University of 
Wisconsin, Vanderbilt University, 
and Yale University.
\end{acknowledgements}

\bibliography{manuscript}

\begin{thebibliography}{}
\expandafter\ifx\csname natexlab\endcsname\relax\def\natexlab#1{#1}\fi
\providecommand{\url}[1]{\href{#1}{#1}}
\providecommand{\dodoi}[1]{doi:~\href{http://doi.org/#1}{\nolinkurl{#1}}}
\providecommand{\doeprint}[1]{\href{http://ascl.net/#1}{\nolinkurl{http://ascl.net/#1}}}
\providecommand{\doarXiv}[1]{\href{https://arxiv.org/abs/#1}{\nolinkurl{https://arxiv.org/abs/#1}}}

\bibitem[{{Blanton} {et~al.}(2017){Blanton}, {Bershady}, {Abolfathi},
  {Albareti}, {Allende Prieto}, {Almeida}, {Alonso-Garc{\'\i}a}, {Anders},
  {Anderson}, {Andrews}, {Aquino-Ort{\'\i}z}, {Arag{\'o}n-Salamanca},
  {Argudo-Fern{\'a}ndez}, {Armengaud}, {Aubourg}, {Avila-Reese}, {Badenes},
  {Bailey}, {Barger}, {Barrera-Ballesteros}, {Bartosz}, {Bates}, {Baumgarten},
  {Bautista}, {Beaton}, {Beers}, {Belfiore}, {Bender}, {Berlind}, {Bernardi},
  {Beutler}, {Bird}, {Bizyaev}, {Blanc}, {Blomqvist}, {Bolton}, {Boquien},
  {Borissova}, {van den Bosch}, {Bovy}, {Brandt}, {Brinkmann}, {Brownstein},
  {Bundy}, {Burgasser}, {Burtin}, {Busca}, {Cappellari}, {Delgado Carigi},
  {Carlberg}, {Carnero Rosell}, {Carrera}, {Chanover}, {Cherinka}, {Cheung},
  {G{\'o}mez Maqueo Chew}, {Chiappini}, {Choi}, {Chojnowski}, {Chuang},
  {Chung}, {Cirolini}, {Clerc}, {Cohen}, {Comparat}, {da Costa}, {Cousinou},
  {Covey}, {Crane}, {Croft}, {Cruz-Gonzalez}, {Garrido Cuadra}, {Cunha},
  {Damke}, {Darling}, {Davies}, {Dawson}, {de la Macorra}, {Dell'Agli}, {De
  Lee}, {Delubac}, {Di Mille}, {Diamond-Stanic}, {Cano-D{\'\i}az}, {Donor},
  {Downes}, {Drory}, {du Mas des Bourboux}, {Duckworth}, {Dwelly}, {Dyer},
  {Ebelke}, {Eigenbrot}, {Eisenstein}, {Emsellem}, {Eracleous}, {Escoffier},
  {Evans}, {Fan}, {Fern{\'a}ndez-Alvar}, {Fernandez-Trincado}, {Feuillet},
  {Finoguenov}, {Fleming}, {Font-Ribera}, {Fredrickson}, {Freischlad},
  {Frinchaboy}, {Fuentes}, {Galbany}, {Garcia-Dias},
  {Garc{\'\i}a-Hern{\'a}ndez}, {Gaulme}, {Geisler}, {Gelfand},
  {Gil-Mar{\'\i}n}, {Gillespie}, {Goddard}, {Gonzalez-Perez}, {Grabowski},
  {Green}, {Grier}, {Gunn}, {Guo}, {Guy}, {Hagen}, {Hahn}, {Hall}, {Harding},
  {Hasselquist}, {Hawley}, {Hearty}, {Gonzalez Hern{\'a}ndez}, {Ho}, {Hogg},
  {Holley-Bockelmann}, {Holtzman}, {Holzer}, {Huehnerhoff}, {Hutchinson},
  {Hwang}, {Ibarra-Medel}, {da Silva Ilha}, {Ivans}, {Ivory}, {Jackson},
  {Jensen}, {Johnson}, {Jones}, {J{\"o}nsson}, {Jullo}, {Kamble}, {Kinemuchi},
  {Kirkby}, {Kitaura}, {Klaene}, {Knapp}, {Kneib}, {Kollmeier}, {Lacerna},
  {Lane}, {Lang}, {Law}, {Lazarz}, {Lee}, {Le Goff}, {Liang}, {Li}, {Li},
  {Lian}, {Lima}, {Lin}, {Lin}, {Bertran de Lis}, {Liu}, {de Icaza Lizaola},
  {Long}, {Lucatello}, {Lundgren}, {MacDonald}, {Deconto Machado}, {MacLeod},
  {Mahadevan}, {Geimba Maia}, {Maiolino}, {Majewski}, {Malanushenko},
  {Malanushenko}, {Manchado}, {Mao}, {Maraston}, {Marques-Chaves}, {Masseron},
  {Masters}, {McBride}, {McDermid}, {McGrath}, {McGreer}, {Medina Pe{\~n}a},
  {Melendez}, {Merloni}, {Merrifield}, {Meszaros}, {Meza}, {Minchev},
  {Minniti}, {Miyaji}, {More}, {Mulchaey}, {M{\"u}ller-S{\'a}nchez}, {Muna},
  {Munoz}, {Myers}, {Nair}, {Nandra}, {Correa do Nascimento}, {Negrete},
  {Ness}, {Newman}, {Nichol}, {Nidever}, {Nitschelm}, {Ntelis}, {O'Connell},
  {Oelkers}, {Oravetz}, {Oravetz}, {Pace}, {Padilla}, {Palanque-Delabrouille},
  {Alonso Palicio}, {Pan}, {Parejko}, {Parikh}, {P{\^a}ris}, {Park}, {Patten},
  {Peirani}, {Pellejero-Ibanez}, {Penny}, {Percival}, {Perez-Fournon},
  {Petitjean}, {Pieri}, {Pinsonneault}, {Pisani}, {Poleski}, {Prada},
  {Prakash}, {Queiroz}, {Raddick}, {Raichoor}, {Barboza Rembold}, {Richstein},
  {Riffel}, {Riffel}, {Rix}, {Robin}, {Rockosi}, {Rodr{\'\i}guez-Torres},
  {Roman-Lopes}, {Rom{\'a}n-Z{\'u}{\~n}iga}, {Rosado}, {Ross}, {Rossi}, {Ruan},
  {Ruggeri}, {Rykoff}, {Salazar-Albornoz}, {Salvato}, {S{\'a}nchez}, {Aguado},
  {S{\'a}nchez-Gallego}, {Santana}, {Santiago}, {Sayres}, {Schiavon}, {da Silva
  Schimoia}, {Schlafly}, {Schlegel}, {Schneider}, {Schultheis}, {Schuster},
  {Schwope}, {Seo}, {Shao}, {Shen}, {Shetrone}, {Shull}, {Simon}, {Skinner},
  {Skrutskie}, {Slosar}, {Smith}, {Sobeck}, {Sobreira}, {Somers}, {Souto},
  {Stark}, {Stassun}, {Stauffer}, {Steinmetz}, {Storchi-Bergmann},
  {Streblyanska}, {Stringfellow}, {Su{\'a}rez}, {Sun}, {Suzuki}, {Szigeti},
  {Taghizadeh-Popp}, {Tang}, {Tao}, {Tayar}, {Tembe}, {Teske}, {Thakar},
  {Thomas}, {Thompson}, {Tinker}, {Tissera}, {Tojeiro}, {Hernandez Toledo}, {de
  la Torre}, {Tremonti}, {Troup}, {Valenzuela}, {Martinez Valpuesta},
  {Vargas-Gonz{\'a}lez}, {Vargas-Maga{\~n}a}, {Vazquez}, {Villanova}, {Vivek},
  {Vogt}, {Wake}, {Walterbos}, {Wang}, {Weaver}, {Weijmans}, {Weinberg},
  {Westfall}, {Whelan}, {Wild}, {Wilson}, {Wood-Vasey}, {Wylezalek}, {Xiao},
  {Yan}, {Yang}, {Ybarra}, {Y{\`e}che}, {Zakamska}, {Zamora}, {Zarrouk},
  {Zasowski}, {Zhang}, {Zhao}, {Zheng}, {Zheng}, {Zhou}, {Zhou}, {Zhu},
  {Zoccali}, \& {Zou}}]{2017AJ....154...28B}
{Blanton}, M.~R., {Bershady}, M.~A., {Abolfathi}, B., {et~al.} 2017, \aj, 154,
  28, \dodoi{10.3847/1538-3881/aa7567}

\bibitem[{Born {et~al.}(1959)Born, Wolf, \& Bhatia}]{born1959principles}
Born, M., Wolf, E., \& Bhatia, A. 1959, Principles of Optics: Electromagnetic
  Theory of Propagation, Interference, and Diffraction of Light (Macmillan).
\newblock \url{https://books.google.com/books?id=QXhKAAAAMAAJ}

\bibitem[{{Colin} {et~al.}(2011){Colin}, {Mohayaee}, {Sarkar}, \&
  {Shafieloo}}]{2011MNRAS.414..264C}
{Colin}, J., {Mohayaee}, R., {Sarkar}, S., \& {Shafieloo}, A. 2011, \mnras,
  414, 264, \dodoi{10.1111/j.1365-2966.2011.18402.x}

\bibitem[{{Gaia Collaboration} {et~al.}(2016){Gaia Collaboration}, {Prusti},
  {de Bruijne}, {Brown}, {Vallenari}, {Babusiaux}, {Bailer-Jones}, {Bastian},
  {Biermann}, {Evans}, {Eyer}, {Jansen}, {Jordi}, {Klioner}, {Lammers},
  {Lindegren}, {Luri}, {Mignard}, {Milligan}, {Panem}, {Poinsignon},
  {Pourbaix}, {Randich}, {Sarri}, {Sartoretti}, {Siddiqui}, {Soubiran},
  {Valette}, {van Leeuwen}, {Walton}, {Aerts}, {Arenou}, {Cropper}, {Drimmel},
  {H{\o}g}, {Katz}, {Lattanzi}, {O'Mullane}, {Grebel}, {Holland}, {Huc},
  {Passot}, {Bramante}, {Cacciari}, {Casta{\~n}eda}, {Chaoul}, {Cheek}, {De
  Angeli}, {Fabricius}, {Guerra}, {Hern{\'a}ndez}, {Jean-Antoine-Piccolo},
  {Masana}, {Messineo}, {Mowlavi}, {Nienartowicz}, {Ord{\'o}{\~n}ez-Blanco},
  {Panuzzo}, {Portell}, {Richards}, {Riello}, {Seabroke}, {Tanga},
  {Th{\'e}venin}, {Torra}, {Els}, {Gracia-Abril}, {Comoretto},
  {Garcia-Reinaldos}, {Lock}, {Mercier}, {Altmann}, {Andrae}, {Astraatmadja},
  {Bellas-Velidis}, {Benson}, {Berthier}, {Blomme}, {Busso}, {Carry},
  {Cellino}, {Clementini}, {Cowell}, {Creevey}, {Cuypers}, {Davidson}, {De
  Ridder}, {de Torres}, {Delchambre}, {Dell'Oro}, {Ducourant}, {Fr{\'e}mat},
  {Garc{\'\i}a-Torres}, {Gosset}, {Halbwachs}, {Hambly}, {Harrison}, {Hauser},
  {Hestroffer}, {Hodgkin}, {Huckle}, {Hutton}, {Jasniewicz}, {Jordan},
  {Kontizas}, {Korn}, {Lanzafame}, {Manteiga}, {Moitinho}, {Muinonen},
  {Osinde}, {Pancino}, {Pauwels}, {Petit}, {Recio-Blanco}, {Robin}, {Sarro},
  {Siopis}, {Smith}, {Smith}, {Sozzetti}, {Thuillot}, {van Reeven}, {Viala},
  {Abbas}, {Abreu Aramburu}, {Accart}, {Aguado}, {Allan}, {Allasia},
  {Altavilla}, {{\'A}lvarez}, {Alves}, {Anderson}, {Andrei}, {Anglada Varela},
  {Antiche}, {Antoja}, {Ant{\'o}n}, {Arcay}, {Atzei}, {Ayache}, {Bach},
  {Baker}, {Balaguer-N{\'u}{\~n}ez}, {Barache}, {Barata}, {Barbier}, {Barblan},
  {Baroni}, {Barrado y Navascu{\'e}s}, {Barros}, {Barstow}, {Becciani},
  {Bellazzini}, {Bellei}, {Bello Garc{\'\i}a}, {Belokurov}, {Bendjoya},
  {Berihuete}, {Bianchi}, {Bienaym{\'e}}, {Billebaud}, {Blagorodnova},
  {Blanco-Cuaresma}, {Boch}, {Bombrun}, {Borrachero}, {Bouquillon}, {Bourda},
  {Bouy}, {Bragaglia}, {Breddels}, {Brouillet}, {Br{\"u}semeister},
  {Bucciarelli}, {Budnik}, {Burgess}, {Burgon}, {Burlacu}, {Busonero}, {Buzzi},
  {Caffau}, {Cambras}, {Campbell}, {Cancelliere}, {Cantat-Gaudin}, {Carlucci},
  {Carrasco}, {Castellani}, {Charlot}, {Charnas}, {Charvet}, {Chassat},
  {Chiavassa}, {Clotet}, {Cocozza}, {Collins}, {Collins}, {Costigan}, {Crifo},
  {Cross}, {Crosta}, {Crowley}, {Dafonte}, {Damerdji}, {Dapergolas}, {David},
  {David}, {De Cat}, {de Felice}, {de Laverny}, {De Luise}, {De March}, {de
  Martino}, {de Souza}, {Debosscher}, {del Pozo}, {Delbo}, {Delgado},
  {Delgado}, {di Marco}, {Di Matteo}, {Diakite}, {Distefano}, {Dolding}, {Dos
  Anjos}, {Drazinos}, {Dur{\'a}n}, {Dzigan}, {Ecale}, {Edvardsson}, {Enke},
  {Erdmann}, {Escolar}, {Espina}, {Evans}, {Eynard Bontemps}, {Fabre},
  {Fabrizio}, {Faigler}, {Falc{\~a}o}, {Farr{\`a}s Casas}, {Faye}, {Federici},
  {Fedorets}, {Fern{\'a}ndez-Hern{\'a}ndez}, {Fernique}, {Fienga}, {Figueras},
  {Filippi}, {Findeisen}, {Fonti}, {Fouesneau}, {Fraile}, {Fraser}, {Fuchs},
  {Furnell}, {Gai}, {Galleti}, {Galluccio}, {Garabato}, {Garc{\'\i}a-Sedano},
  {Gar{\'e}}, {Garofalo}, {Garralda}, {Gavras}, {Gerssen}, {Geyer}, {Gilmore},
  {Girona}, {Giuffrida}, {Gomes}, {Gonz{\'a}lez-Marcos},
  {Gonz{\'a}lez-N{\'u}{\~n}ez}, {Gonz{\'a}lez-Vidal}, {Granvik}, {Guerrier},
  {Guillout}, {Guiraud}, {G{\'u}rpide}, {Guti{\'e}rrez-S{\'a}nchez}, {Guy},
  {Haigron}, {Hatzidimitriou}, {Haywood}, {Heiter}, {Helmi}, {Hobbs},
  {Hofmann}, {Holl}, {Holland}, {Hunt}, {Hypki}, {Icardi}, {Irwin}, {Jevardat
  de Fombelle}, {Jofr{\'e}}, {Jonker}, {Jorissen}, {Julbe}, {Karampelas},
  {Kochoska}, {Kohley}, {Kolenberg}, {Kontizas}, {Koposov}, {Kordopatis},
  {Koubsky}, {Kowalczyk}, {Krone-Martins}, {Kudryashova}, {Kull}, {Bachchan},
  {Lacoste-Seris}, {Lanza}, {Lavigne}, {Le Poncin-Lafitte}, {Lebreton},
  {Lebzelter}, {Leccia}, {Leclerc}, {Lecoeur-Taibi}, {Lemaitre}, {Lenhardt},
  {Leroux}, {Liao}, {Licata}, {Lindstr{\o}m}, {Lister}, {Livanou}, {Lobel},
  {L{\"o}ffler}, {L{\'o}pez}, {Lopez-Lozano}, {Lorenz}, {Loureiro},
  {MacDonald}, {Magalh{\~a}es Fernandes}, {Managau}, {Mann}, {Mantelet},
  {Marchal}, {Marchant}, {Marconi}, {Marie}, {Marinoni}, {Marrese},
  {Marschalk{\'o}}, {Marshall}, {Mart{\'\i}n-Fleitas}, {Martino}, {Mary},
  {Matijevi{\v{c}}}, {Mazeh}, {McMillan}, {Messina}, {Mestre}, {Michalik},
  {Millar}, {Miranda}, {Molina}, {Molinaro}, {Molinaro}, {Moln{\'a}r},
  {Moniez}, {Montegriffo}, {Monteiro}, {Mor}, {Mora}, {Morbidelli}, {Morel},
  {Morgenthaler}, {Morley}, {Morris}, {Mulone}, {Muraveva}, {Musella},
  {Narbonne}, {Nelemans}, {Nicastro}, {Noval}, {Ord{\'e}novic},
  {Ordieres-Mer{\'e}}, {Osborne}, {Pagani}, {Pagano}, {Pailler}, {Palacin},
  {Palaversa}, {Parsons}, {Paulsen}, {Pecoraro}, {Pedrosa}, {Pentik{\"a}inen},
  {Pereira}, {Pichon}, {Piersimoni}, {Pineau}, {Plachy}, {Plum}, {Poujoulet},
  {Pr{\v{s}}a}, {Pulone}, {Ragaini}, {Rago}, {Rambaux}, {Ramos-Lerate},
  {Ranalli}, {Rauw}, {Read}, {Regibo}, {Renk}, {Reyl{\'e}}, {Ribeiro},
  {Rimoldini}, {Ripepi}, {Riva}, {Rixon}, {Roelens}, {Romero-G{\'o}mez},
  {Rowell}, {Royer}, {Rudolph}, {Ruiz-Dern}, {Sadowski}, {Sagrist{\`a}
  Sell{\'e}s}, {Sahlmann}, {Salgado}, {Salguero}, {Sarasso}, {Savietto},
  {Schnorhk}, {Schultheis}, {Sciacca}, {Segol}, {Segovia}, {Segransan},
  {Serpell}, {Shih}, {Smareglia}, {Smart}, {Smith}, {Solano}, {Solitro},
  {Sordo}, {Soria Nieto}, {Souchay}, {Spagna}, {Spoto}, {Stampa}, {Steele},
  {Steidelm{\"u}ller}, {Stephenson}, {Stoev}, {Suess}, {S{\"u}veges}, {Surdej},
  {Szabados}, {Szegedi-Elek}, {Tapiador}, {Taris}, {Tauran}, {Taylor},
  {Teixeira}, {Terrett}, {Tingley}, {Trager}, {Turon}, {Ulla}, {Utrilla},
  {Valentini}, {van Elteren}, {Van Hemelryck}, {van Leeuwen}, {Varadi},
  {Vecchiato}, {Veljanoski}, {Via}, {Vicente}, {Vogt}, {Voss}, {Votruba},
  {Voutsinas}, {Walmsley}, {Weiler}, {Weingrill}, {Werner}, {Wevers},
  {Whitehead}, {Wyrzykowski}, {Yoldas}, {{\v{Z}}erjal}, {Zucker}, {Zurbach},
  {Zwitter}, {Alecu}, {Allen}, {Allende Prieto}, {Amorim},
  {Anglada-Escud{\'e}}, {Arsenijevic}, {Azaz}, {Balm}, {Beck}, {Bernstein},
  {Bigot}, {Bijaoui}, {Blasco}, {Bonfigli}, {Bono}, {Boudreault}, {Bressan},
  {Brown}, {Brunet}, {Bunclark}, {Buonanno}, {Butkevich}, {Carret}, {Carrion},
  {Chemin}, {Ch{\'e}reau}, {Corcione}, {Darmigny}, {de Boer}, {de Teodoro}, {de
  Zeeuw}, {Delle Luche}, {Domingues}, {Dubath}, {Fodor}, {Fr{\'e}zouls},
  {Fries}, {Fustes}, {Fyfe}, {Gallardo}, {Gallegos}, {Gardiol}, {Gebran},
  {Gomboc}, {G{\'o}mez}, {Grux}, {Gueguen}, {Heyrovsky}, {Hoar}, {Iannicola},
  {Isasi Parache}, {Janotto}, {Joliet}, {Jonckheere}, {Keil}, {Kim},
  {Klagyivik}, {Klar}, {Knude}, {Kochukhov}, {Kolka}, {Kos}, {Kutka}, {Lainey},
  {LeBouquin}, {Liu}, {Loreggia}, {Makarov}, {Marseille}, {Martayan},
  {Martinez-Rubi}, {Massart}, {Meynadier}, {Mignot}, {Munari}, {Nguyen},
  {Nordlander}, {Ocvirk}, {O'Flaherty}, {Olias Sanz}, {Ortiz}, {Osorio},
  {Oszkiewicz}, {Ouzounis}, {Palmer}, {Park}, {Pasquato}, {Peltzer}, {Peralta},
  {P{\'e}turaud}, {Pieniluoma}, {Pigozzi}, {Poels}, {Prat}, {Prod'homme},
  {Raison}, {Rebordao}, {Risquez}, {Rocca-Volmerange}, {Rosen}, {Ruiz-Fuertes},
  {Russo}, {Sembay}, {Serraller Vizcaino}, {Short}, {Siebert}, {Silva},
  {Sinachopoulos}, {Slezak}, {Soffel}, {Sosnowska}, {Strai{\v{z}}ys}, {ter
  Linden}, {Terrell}, {Theil}, {Tiede}, {Troisi}, {Tsalmantza}, {Tur},
  {Vaccari}, {Vachier}, {Valles}, {Van Hamme}, {Veltz}, {Virtanen}, {Wallut},
  {Wichmann}, {Wilkinson}, {Ziaeepour}, \& {Zschocke}}]{2016A&A...595A...1G}
{Gaia Collaboration}, {Prusti}, T., {de Bruijne}, J.~H.~J., {et~al.} 2016,
  \aap, 595, A1, \dodoi{10.1051/0004-6361/201629272}

\bibitem[{{Gaia Collaboration} {et~al.}(2018){Gaia Collaboration}, {Mignard},
  {Klioner}, {Lindegren}, {Hern{\'a}ndez}, {Bastian}, {Bombrun}, {Hobbs},
  {Lammers}, {Michalik}, {Ramos-Lerate}, {Biermann},
  {Fern{\'a}ndez-Hern{\'a}ndez}, {Geyer}, {Hilger}, {Siddiqui},
  {Steidelm{\"u}ller}, {Babusiaux}, {Barache}, {Lambert}, {Andrei}, {Bourda},
  {Charlot}, {Brown}, {Vallenari}, {Prusti}, {de Bruijne}, {Bailer-Jones},
  {Evans}, {Eyer}, {Jansen}, {Jordi}, {Luri}, {Panem}, {Pourbaix}, {Randich},
  {Sartoretti}, {Soubiran}, {van Leeuwen}, {Walton}, {Arenou}, {Cropper},
  {Drimmel}, {Katz}, {Lattanzi}, {Bakker}, {Cacciari}, {Casta{\~n}eda},
  {Chaoul}, {Cheek}, {De Angeli}, {Fabricius}, {Guerra}, {Holl}, {Masana},
  {Messineo}, {Mowlavi}, {Nienartowicz}, {Panuzzo}, {Portell}, {Riello},
  {Seabroke}, {Tanga}, {Th{\'e}venin}, {Gracia-Abril}, {Comoretto},
  {Garcia-Reinaldos}, {Teyssier}, {Altmann}, {Andrae}, {Audard},
  {Bellas-Velidis}, {Benson}, {Berthier}, {Blomme}, {Burgess}, {Busso},
  {Carry}, {Cellino}, {Clementini}, {Clotet}, {Creevey}, {Davidson}, {De
  Ridder}, {Delchambre}, {Dell'Oro}, {Ducourant}, {Fouesneau}, {Fr{\'e}mat},
  {Galluccio}, {Garc{\'\i}a-Torres}, {Gonz{\'a}lez-N{\'u}{\~n}ez},
  {Gonz{\'a}lez-Vidal}, {Gosset}, {Guy}, {Halbwachs}, {Hambly}, {Harrison},
  {Hestroffer}, {Hodgkin}, {Hutton}, {Jasniewicz}, {Jean-Antoine-Piccolo},
  {Jordan}, {Korn}, {Krone-Martins}, {Lanzafame}, {Lebzelter}, {L{\"o}ffler},
  {Manteiga}, {Marrese}, {Mart{\'\i}n-Fleitas}, {Moitinho}, {Mora}, {Muinonen},
  {Osinde}, {Pancino}, {Pauwels}, {Petit}, {Recio-Blanco}, {Richards},
  {Rimoldini}, {Robin}, {Sarro}, {Siopis}, {Smith}, {Sozzetti}, {S{\"u}veges},
  {Torra}, {van Reeven}, {Abbas}, {Abreu Aramburu}, {Accart}, {Aerts},
  {Altavilla}, {{\'A}lvarez}, {Alvarez}, {Alves}, {Anderson}, {Anglada Varela},
  {Antiche}, {Antoja}, {Arcay}, {Astraatmadja}, {Bach}, {Baker},
  {Balaguer-N{\'u}{\~n}ez}, {Balm}, {Barata}, {Barbato}, {Barblan}, {Barklem},
  {Barrado}, {Barros}, {Barstow}, {Bartholom{\'e} Mu{\~n}oz}, {Bassilana},
  {Becciani}, {Bellazzini}, {Berihuete}, {Bertone}, {Bianchi}, {Bienaym{\'e}},
  {Blanco-Cuaresma}, {Boch}, {Boeche}, {Borrachero}, {Bossini}, {Bouquillon},
  {Bragaglia}, {Bramante}, {Breddels}, {Bressan}, {Brouillet},
  {Br{\"u}semeister}, {Brugaletta}, {Bucciarelli}, {Burlacu}, {Busonero},
  {Butkevich}, {Buzzi}, {Caffau}, {Cancelliere}, {Cannizzaro}, {Cantat-Gaudin},
  {Carballo}, {Carlucci}, {Carrasco}, {Casamiquela}, {Castellani},
  {Castro-Ginard}, {Chemin}, {Chiavassa}, {Cocozza}, {Costigan}, {Cowell},
  {Crifo}, {Crosta}, {Crowley}, {Cuypers}, {Dafonte}, {Damerdji}, {Dapergolas},
  {David}, {David}, {de Laverny}, {De Luise}, {De March}, {de Souza}, {de
  Torres}, {Debosscher}, {del Pozo}, {Delbo}, {Delgado}, {Delgado}, {Diakite},
  {Diener}, {Distefano}, {Dolding}, {Drazinos}, {Dur{\'a}n}, {Edvardsson},
  {Enke}, {Eriksson}, {Esquej}, {Eynard Bontemps}, {Fabre}, {Fabrizio},
  {Faigler}, {Falc{\~a}o}, {Farr{\`a}s Casas}, {Federici}, {Fedorets},
  {Fernique}, {Figueras}, {Filippi}, {Findeisen}, {Fonti}, {Fraile}, {Fraser},
  {Fr{\'e}zouls}, {Gai}, {Galleti}, {Garabato}, {Garc{\'\i}a-Sedano},
  {Garofalo}, {Garralda}, {Gavel}, {Gavras}, {Gerssen}, {Giacobbe}, {Gilmore},
  {Girona}, {Giuffrida}, {Glass}, {Gomes}, {Granvik}, {Gueguen}, {Guerrier},
  {Guiraud}, {Guti{\'e}}, {Haigron}, {Hatzidimitriou}, {Hauser}, {Haywood},
  {Heiter}, {Helmi}, {Heu}, {Hofmann}, {Holland}, {Huckle}, {Hypki}, {Icardi},
  {Jan{\ss}en}, {Jevardat de Fombelle}, {Jonker}, {Juh{\'a}sz}, {Julbe},
  {Karampelas}, {Kewley}, {Klar}, {Kochoska}, {Kohley}, {Kolenberg},
  {Kontizas}, {Kontizas}, {Koposov}, {Kordopatis}, {Kostrzewa-Rutkowska},
  {Koubsky}, {Lanza}, {Lasne}, {Lavigne}, {Le Fustec}, {Le Poncin-Lafitte},
  {Lebreton}, {Leccia}, {Leclerc}, {Lecoeur-Taibi}, {Lenhardt}, {Leroux},
  {Liao}, {Licata}, {Lindstr{\o}m}, {Lister}, {Livanou}, {Lobel}, {L{\'o}pez},
  {Managau}, {Mann}, {Mantelet}, {Marchal}, {Marchant}, {Marconi}, {Marinoni},
  {Marschalk{\'o}}, {Marshall}, {Martino}, {Marton}, {Mary}, {Massari},
  {Matijevi{\v{c}}}, {Mazeh}, {McMillan}, {Messina}, {Millar}, {Molina},
  {Molinaro}, {Moln{\'a}r}, {Montegriffo}, {Mor}, {Morbidelli}, {Morel},
  {Morris}, {Mulone}, {Muraveva}, {Musella}, {Nelemans}, {Nicastro}, {Noval},
  {O'Mullane}, {Ord{\'e}novic}, {Ord{\'o}{\~n}ez-Blanco}, {Osborne}, {Pagani},
  {Pagano}, {Pailler}, {Palacin}, {Palaversa}, {Panahi}, {Pawlak},
  {Piersimoni}, {Pineau}, {Plachy}, {Plum}, {Poggio}, {Poujoulet},
  {Pr{\v{s}}a}, {Pulone}, {Racero}, {Ragaini}, {Rambaux}, {Regibo},
  {Reyl{\'e}}, {Riclet}, {Ripepi}, {Riva}, {Rivard}, {Rixon}, {Roegiers},
  {Roelens}, {Romero-G{\'o}mez}, {Rowell}, {Royer}, {Ruiz-Dern}, {Sadowski},
  {Sagrist{\`a} Sell{\'e}s}, {Sahlmann}, {Salgado}, {Salguero}, {Sanna},
  {Santana-Ros}, {Sarasso}, {Savietto}, {Schultheis}, {Sciacca}, {Segol},
  {Segovia}, {S{\'e}gransan}, {Shih}, {Siltala}, {Silva}, {Smart}, {Smith},
  {Solano}, {Solitro}, {Sordo}, {Soria Nieto}, {Souchay}, {Spagna}, {Spoto},
  {Stampa}, {Steele}, {Stephenson}, {Stoev}, {Suess}, {Surdej}, {Szabados},
  {Szegedi-Elek}, {Tapiador}, {Taris}, {Tauran}, {Taylor}, {Teixeira},
  {Terrett}, {Teyssandier}, {Thuillot}, {Titarenko}, {Torra Clotet}, {Turon},
  {Ulla}, {Utrilla}, {Uzzi}, {Vaillant}, {Valentini}, {Valette}, {van Elteren},
  {Van Hemelryck}, {van Leeuwen}, {Vaschetto}, {Vecchiato}, {Veljanoski},
  {Viala}, {Vicente}, {Vogt}, {von Essen}, {Voss}, {Votruba}, {Voutsinas},
  {Walmsley}, {Weiler}, {Wertz}, {Wevers}, {Wyrzykowski}, {Yoldas},
  {{\v{Z}}erjal}, {Ziaeepour}, {Zorec}, {Zschocke}, {Zucker}, {Zurbach}, \&
  {Zwitter}}]{2018A&A...616A..14G}
{Gaia Collaboration}, {Mignard}, F., {Klioner}, S.~A., {et~al.} 2018, \aap,
  616, A14, \dodoi{10.1051/0004-6361/201832916}

\bibitem[{{Gaia Collaboration} {et~al.}(2021{\natexlab{a}}){Gaia
  Collaboration}, {Brown}, {Vallenari}, {Prusti}, {de Bruijne}, {Babusiaux},
  {Biermann}, {Creevey}, {Evans}, {Eyer}, {Hutton}, {Jansen}, {Jordi},
  {Klioner}, {Lammers}, {Lindegren}, {Luri}, {Mignard}, {Panem}, {Pourbaix},
  {Randich}, {Sartoretti}, {Soubiran}, {Walton}, {Arenou}, {Bailer-Jones},
  {Bastian}, {Cropper}, {Drimmel}, {Katz}, {Lattanzi}, {van Leeuwen}, {Bakker},
  {Cacciari}, {Casta{\~n}eda}, {De Angeli}, {Ducourant}, {Fabricius},
  {Fouesneau}, {Fr{\'e}mat}, {Guerra}, {Guerrier}, {Guiraud}, {Jean-Antoine
  Piccolo}, {Masana}, {Messineo}, {Mowlavi}, {Nicolas}, {Nienartowicz},
  {Pailler}, {Panuzzo}, {Riclet}, {Roux}, {Seabroke}, {Sordo}, {Tanga},
  {Th{\'e}venin}, {Gracia-Abril}, {Portell}, {Teyssier}, {Altmann}, {Andrae},
  {Bellas-Velidis}, {Benson}, {Berthier}, {Blomme}, {Brugaletta}, {Burgess},
  {Busso}, {Carry}, {Cellino}, {Cheek}, {Clementini}, {Damerdji}, {Davidson},
  {Delchambre}, {Dell'Oro}, {Fern{\'a}ndez-Hern{\'a}ndez}, {Galluccio},
  {Garc{\'\i}a-Lario}, {Garcia-Reinaldos}, {Gonz{\'a}lez-N{\'u}{\~n}ez},
  {Gosset}, {Haigron}, {Halbwachs}, {Hambly}, {Harrison}, {Hatzidimitriou},
  {Heiter}, {Hern{\'a}ndez}, {Hestroffer}, {Hodgkin}, {Holl}, {Jan{\ss}en},
  {Jevardat de Fombelle}, {Jordan}, {Krone-Martins}, {Lanzafame},
  {L{\"o}ffler}, {Lorca}, {Manteiga}, {Marchal}, {Marrese}, {Moitinho}, {Mora},
  {Muinonen}, {Osborne}, {Pancino}, {Pauwels}, {Petit}, {Recio-Blanco},
  {Richards}, {Riello}, {Rimoldini}, {Robin}, {Roegiers}, {Rybizki}, {Sarro},
  {Siopis}, {Smith}, {Sozzetti}, {Ulla}, {Utrilla}, {van Leeuwen}, {van
  Reeven}, {Abbas}, {Abreu Aramburu}, {Accart}, {Aerts}, {Aguado}, {Ajaj},
  {Altavilla}, {{\'A}lvarez}, {{\'A}lvarez Cid-Fuentes}, {Alves}, {Anderson},
  {Anglada Varela}, {Antoja}, {Audard}, {Baines}, {Baker},
  {Balaguer-N{\'u}{\~n}ez}, {Balbinot}, {Balog}, {Barache}, {Barbato},
  {Barros}, {Barstow}, {Bartolom{\'e}}, {Bassilana}, {Bauchet},
  {Baudesson-Stella}, {Becciani}, {Bellazzini}, {Bernet}, {Bertone}, {Bianchi},
  {Blanco-Cuaresma}, {Boch}, {Bombrun}, {Bossini}, {Bouquillon}, {Bragaglia},
  {Bramante}, {Breedt}, {Bressan}, {Brouillet}, {Bucciarelli}, {Burlacu},
  {Busonero}, {Butkevich}, {Buzzi}, {Caffau}, {Cancelliere}, {C{\'a}novas},
  {Cantat-Gaudin}, {Carballo}, {Carlucci}, {Carnerero}, {Carrasco},
  {Casamiquela}, {Castellani}, {Castro-Ginard}, {Castro Sampol}, {Chaoul},
  {Charlot}, {Chemin}, {Chiavassa}, {Cioni}, {Comoretto}, {Cooper}, {Cornez},
  {Cowell}, {Crifo}, {Crosta}, {Crowley}, {Dafonte}, {Dapergolas}, {David},
  {David}, {de Laverny}, {De Luise}, {De March}, {De Ridder}, {de Souza}, {de
  Teodoro}, {de Torres}, {del Peloso}, {del Pozo}, {Delbo}, {Delgado},
  {Delgado}, {Delisle}, {Di Matteo}, {Diakite}, {Diener}, {Distefano},
  {Dolding}, {Eappachen}, {Edvardsson}, {Enke}, {Esquej}, {Fabre}, {Fabrizio},
  {Faigler}, {Fedorets}, {Fernique}, {Fienga}, {Figueras}, {Fouron},
  {Fragkoudi}, {Fraile}, {Franke}, {Gai}, {Garabato}, {Garcia-Gutierrez},
  {Garc{\'\i}a-Torres}, {Garofalo}, {Gavras}, {Gerlach}, {Geyer}, {Giacobbe},
  {Gilmore}, {Girona}, {Giuffrida}, {Gomel}, {Gomez}, {Gonzalez-Santamaria},
  {Gonz{\'a}lez-Vidal}, {Granvik}, {Guti{\'e}rrez-S{\'a}nchez}, {Guy},
  {Hauser}, {Haywood}, {Helmi}, {Hidalgo}, {Hilger}, {H{\l}adczuk}, {Hobbs},
  {Holland}, {Huckle}, {Jasniewicz}, {Jonker}, {Juaristi Campillo}, {Julbe},
  {Karbevska}, {Kervella}, {Khanna}, {Kochoska}, {Kontizas}, {Kordopatis},
  {Korn}, {Kostrzewa-Rutkowska}, {Kruszy{\'n}ska}, {Lambert}, {Lanza}, {Lasne},
  {Le Campion}, {Le Fustec}, {Lebreton}, {Lebzelter}, {Leccia}, {Leclerc},
  {Lecoeur-Taibi}, {Liao}, {Licata}, {Lindstr{\o}m}, {Lister}, {Livanou},
  {Lobel}, {Madrero Pardo}, {Managau}, {Mann}, {Marchant}, {Marconi}, {Marcos
  Santos}, {Marinoni}, {Marocco}, {Marshall}, {Martin Polo},
  {Mart{\'\i}n-Fleitas}, {Masip}, {Massari}, {Mastrobuono-Battisti}, {Mazeh},
  {McMillan}, {Messina}, {Michalik}, {Millar}, {Mints}, {Molina}, {Molinaro},
  {Moln{\'a}r}, {Montegriffo}, {Mor}, {Morbidelli}, {Morel}, {Morris},
  {Mulone}, {Munoz}, {Muraveva}, {Murphy}, {Musella}, {Noval}, {Ord{\'e}novic},
  {Orr{\`u}}, {Osinde}, {Pagani}, {Pagano}, {Palaversa}, {Palicio}, {Panahi},
  {Pawlak}, {Pe{\~n}alosa Esteller}, {Penttil{\"a}}, {Piersimoni}, {Pineau},
  {Plachy}, {Plum}, {Poggio}, {Poretti}, {Poujoulet}, {Pr{\v{s}}a}, {Pulone},
  {Racero}, {Ragaini}, {Rainer}, {Raiteri}, {Rambaux}, {Ramos}, {Ramos-Lerate},
  {Re Fiorentin}, {Regibo}, {Reyl{\'e}}, {Ripepi}, {Riva}, {Rixon}, {Robichon},
  {Robin}, {Roelens}, {Rohrbasser}, {Romero-G{\'o}mez}, {Rowell}, {Royer},
  {Rybicki}, {Sadowski}, {Sagrist{\`a} Sell{\'e}s}, {Sahlmann}, {Salgado},
  {Salguero}, {Samaras}, {Sanchez Gimenez}, {Sanna}, {Santove{\~n}a},
  {Sarasso}, {Schultheis}, {Sciacca}, {Segol}, {Segovia}, {S{\'e}gransan},
  {Semeux}, {Shahaf}, {Siddiqui}, {Siebert}, {Siltala}, {Slezak}, {Smart},
  {Solano}, {Solitro}, {Souami}, {Souchay}, {Spagna}, {Spoto}, {Steele},
  {Steidelm{\"u}ller}, {Stephenson}, {S{\"u}veges}, {Szabados}, {Szegedi-Elek},
  {Taris}, {Tauran}, {Taylor}, {Teixeira}, {Thuillot}, {Tonello}, {Torra},
  {Torra}, {Turon}, {Unger}, {Vaillant}, {van Dillen}, {Vanel}, {Vecchiato},
  {Viala}, {Vicente}, {Voutsinas}, {Weiler}, {Wevers}, {Wyrzykowski}, {Yoldas},
  {Yvard}, {Zhao}, {Zorec}, {Zucker}, {Zurbach}, \&
  {Zwitter}}]{2021A&A...649A...1G}
{Gaia Collaboration}, {Brown}, A.~G.~A., {Vallenari}, A., {et~al.}
  2021{\natexlab{a}}, \aap, 649, A1, \dodoi{10.1051/0004-6361/202039657}

\bibitem[{{Gaia Collaboration} {et~al.}(2021{\natexlab{b}}){Gaia
  Collaboration}, {Klioner}, {Mignard}, {Lindegren}, {Bastian}, {McMillan},
  {Hern{\'a}ndez}, {Hobbs}, {Ramos-Lerate}, {Biermann}, {Bombrun}, {de Torres},
  {Gerlach}, {Geyer}, {Hilger}, {Lammers}, {Steidelm{\"u}ller}, {Stephenson},
  {Brown}, {Vallenari}, {Prusti}, {de Bruijne}, {Babusiaux}, {Creevey},
  {Evans}, {Eyer}, {Hutton}, {Jansen}, {Jordi}, {Luri}, {Panem}, {Pourbaix},
  {Randich}, {Sartoretti}, {Soubiran}, {Walton}, {Arenou}, {Bailer-Jones},
  {Cropper}, {Drimmel}, {Katz}, {Lattanzi}, {van Leeuwen}, {Bakker},
  {Casta{\~n}eda}, {De Angeli}, {Ducourant}, {Fabricius}, {Fouesneau},
  {Fr{\'e}mat}, {Guerra}, {Guerrier}, {Guiraud}, {Jean-Antoine Piccolo},
  {Masana}, {Messineo}, {Mowlavi}, {Nicolas}, {Nienartowicz}, {Pailler},
  {Panuzzo}, {Riclet}, {Roux}, {Seabroke}, {Sordo}, {Tanga}, {Th{\'e}venin},
  {Gracia-Abril}, {Portell}, {Teyssier}, {Altmann}, {Andrae}, {Bellas-Velidis},
  {Benson}, {Berthier}, {Blomme}, {Brugaletta}, {Burgess}, {Busso}, {Carry},
  {Cellino}, {Cheek}, {Clementini}, {Damerdji}, {Davidson}, {Delchambre},
  {Dell'Oro}, {Fern{\'a}ndez-Hern{\'a}ndez}, {Galluccio}, {Garc{\'\i}a-Lario},
  {Garcia-Reinaldos}, {Gonz{\'a}lez-N{\'u}{\~n}ez}, {Gosset}, {Haigron},
  {Halbwachs}, {Hambly}, {Harrison}, {Hatzidimitriou}, {Heiter}, {Hestroffer},
  {Hodgkin}, {Holl}, {Jan{\ss}en}, {Jevardat de Fombelle}, {Jordan},
  {Krone-Martins}, {Lanzafame}, {L{\"o}ffler}, {Lorca}, {Manteiga}, {Marchal},
  {Marrese}, {Moitinho}, {Mora}, {Muinonen}, {Osborne}, {Pancino}, {Pauwels},
  {Recio-Blanco}, {Richards}, {Riello}, {Rimoldini}, {Robin}, {Roegiers},
  {Rybizki}, {Sarro}, {Siopis}, {Smith}, {Sozzetti}, {Ulla}, {Utrilla}, {van
  Leeuwen}, {van Reeven}, {Abbas}, {Abreu Aramburu}, {Accart}, {Aerts},
  {Aguado}, {Ajaj}, {Altavilla}, {{\'A}lvarez}, {{\'A}lvarez Cid-Fuentes},
  {Alves}, {Anderson}, {Anglada Varela}, {Antoja}, {Audard}, {Baines}, {Baker},
  {Balaguer-N{\'u}{\~n}ez}, {Balbinot}, {Balog}, {Barache}, {Barbato},
  {Barros}, {Barstow}, {Bartolom{\'e}}, {Bassilana}, {Bauchet},
  {Baudesson-Stella}, {Becciani}, {Bellazzini}, {Bernet}, {Bertone}, {Bianchi},
  {Blanco-Cuaresma}, {Boch}, {Bossini}, {Bouquillon}, {Bramante}, {Breedt},
  {Bressan}, {Brouillet}, {Bucciarelli}, {Burlacu}, {Busonero}, {Butkevich},
  {Buzzi}, {Caffau}, {Cancelliere}, {C{\'a}novas}, {Cantat-Gaudin}, {Carballo},
  {Carlucci}, {Carnerero}, {Carrasco}, {Casamiquela}, {Castellani},
  {Castro-Ginard}, {Castro Sampol}, {Chaoul}, {Charlot}, {Chemin}, {Chiavassa},
  {Comoretto}, {Cooper}, {Cornez}, {Cowell}, {Crifo}, {Crosta}, {Crowley},
  {Dafonte}, {Dapergolas}, {David}, {David}, {de Laverny}, {De Luise}, {De
  March}, {De Ridder}, {de Souza}, {de Teodoro}, {del Peloso}, {del Pozo},
  {Delgado}, {Delgado}, {Delisle}, {Di Matteo}, {Diakite}, {Diener},
  {Distefano}, {Dolding}, {Eappachen}, {Enke}, {Esquej}, {Fabre}, {Fabrizio},
  {Faigler}, {Fedorets}, {Fernique}, {Fienga}, {Figueras}, {Fouron},
  {Fragkoudi}, {Fraile}, {Franke}, {Gai}, {Garabato}, {Garcia-Gutierrez},
  {Garc{\'\i}a-Torres}, {Garofalo}, {Gavras}, {Giacobbe}, {Gilmore}, {Girona},
  {Giuffrida}, {Gomez}, {Gonzalez-Santamaria}, {Gonz{\'a}lez-Vidal}, {Granvik},
  {Guti{\'e}rrez-S{\'a}nchez}, {Guy}, {Hauser}, {Haywood}, {Helmi}, {Hidalgo},
  {H{\l}adczuk}, {Holland}, {Huckle}, {Jasniewicz}, {Jonker}, {Juaristi
  Campillo}, {Julbe}, {Karbevska}, {Kervella}, {Khanna}, {Kochoska},
  {Kordopatis}, {Korn}, {Kostrzewa-Rutkowska}, {Kruszy{\'n}ska}, {Lambert},
  {Lanza}, {Lasne}, {Le Campion}, {Le Fustec}, {Lebreton}, {Lebzelter},
  {Leccia}, {Leclerc}, {Lecoeur-Taibi}, {Liao}, {Licata}, {Lindstr{\o}m},
  {Lister}, {Livanou}, {Lobel}, {Madrero Pardo}, {Managau}, {Mann}, {Marchant},
  {Marconi}, {Marcos Santos}, {Marinoni}, {Marocco}, {Marshall}, {Martin Polo},
  {Mart{\'\i}n-Fleitas}, {Masip}, {Massari}, {Mastrobuono-Battisti}, {Mazeh},
  {Messina}, {Michalik}, {Millar}, {Mints}, {Molina}, {Molinaro}, {Moln{\'a}r},
  {Montegriffo}, {Mor}, {Morbidelli}, {Morel}, {Morris}, {Mulone}, {Munoz},
  {Muraveva}, {Murphy}, {Musella}, {Noval}, {Ord{\'e}novic}, {Orr{\`u}},
  {Osinde}, {Pagani}, {Pagano}, {Palaversa}, {Palicio}, {Panahi}, {Pawlak},
  {Pe{\~n}alosa Esteller}, {Penttil{\"a}}, {Piersimoni}, {Pineau}, {Plachy},
  {Plum}, {Poggio}, {Poretti}, {Poujoulet}, {Pr{\v{s}}a}, {Pulone}, {Racero},
  {Ragaini}, {Rainer}, {Raiteri}, {Rambaux}, {Ramos}, {Re Fiorentin}, {Regibo},
  {Reyl{\'e}}, {Ripepi}, {Riva}, {Rixon}, {Robichon}, {Robin}, {Roelens},
  {Rohrbasser}, {Romero-G{\'o}mez}, {Rowell}, {Royer}, {Rybicki}, {Sadowski},
  {Sagrist{\`a} Sell{\'e}s}, {Sahlmann}, {Salgado}, {Salguero}, {Samaras},
  {Sanchez Gimenez}, {Sanna}, {Santove{\~n}a}, {Sarasso}, {Schultheis},
  {Sciacca}, {Segol}, {Segovia}, {S{\'e}gransan}, {Semeux}, {Siddiqui},
  {Siebert}, {Siltala}, {Slezak}, {Smart}, {Solano}, {Solitro}, {Souami},
  {Souchay}, {Spagna}, {Spoto}, {Steele}, {S{\"u}veges}, {Szabados},
  {Szegedi-Elek}, {Taris}, {Tauran}, {Taylor}, {Teixeira}, {Thuillot},
  {Tonello}, {Torra}, {Torra}, {Turon}, {Unger}, {Vaillant}, {van Dillen},
  {Vanel}, {Vecchiato}, {Viala}, {Vicente}, {Voutsinas}, {Weiler}, {Wevers},
  {Wyrzykowski}, {Yoldas}, {Yvard}, {Zhao}, {Zorec}, {Zucker}, {Zurbach}, \&
  {Zwitter}}]{2021A&A...649A...9G}
{Gaia Collaboration}, {Klioner}, S.~A., {Mignard}, F., {et~al.}
  2021{\natexlab{b}}, \aap, 649, A9, \dodoi{10.1051/0004-6361/202039734}

\bibitem[{{Hobbs} {et~al.}(2021){Hobbs}, {Brown}, {H{\o}g}, {Jordi}, {Kawata},
  {Tanga}, {Klioner}, {Sozzetti}, {Wyrzykowski}, {Walton}, {Vallenari},
  {Makarov}, {Rybizki}, {Jim{\'e}nez-Esteban}, {Caballero}, {McMillan},
  {Secrest}, {Mor}, {Andrews}, {Zwitter}, {Chiappini}, {Fynbo}, {Ting},
  {Hestroffer}, {Lindegren}, {McArthur}, {Gouda}, {Moore}, {Gonzalez}, \&
  {Vaccari}}]{2021ExA....51..783H}
{Hobbs}, D., {Brown}, A., {H{\o}g}, E., {et~al.} 2021, Experimental Astronomy,
  51, 783, \dodoi{10.1007/s10686-021-09705-z}

\bibitem[{{H{\o}g}(2021)}]{2021arXiv210707177H}
{H{\o}g}, E. 2021, arXiv e-prints, arXiv:2107.07177.
\newblock \doarXiv{2107.07177}

\bibitem[{{Kopeikin} \& {Makarov}(2021)}]{2021FrASS...8....9K}
{Kopeikin}, S.~M., \& {Makarov}, V.~V. 2021, Frontiers in Astronomy and Space
  Sciences, 8, 9, \dodoi{10.3389/fspas.2021.639706}

\bibitem[{{Kovalev} {et~al.}(2017){Kovalev}, {Petrov}, \&
  {Plavin}}]{2017A&A...598L...1K}
{Kovalev}, Y.~Y., {Petrov}, L., \& {Plavin}, A.~V. 2017, \aap, 598, L1,
  \dodoi{10.1051/0004-6361/201630031}

\bibitem[{{Makarov}(2021)}]{2021AJ....161..289M}
{Makarov}, V.~V. 2021, \aj, 161, 289, \dodoi{10.3847/1538-3881/abf249}

\bibitem[{{Makarov} {et~al.}(2012){Makarov}, {Veillette}, {Hennessy}, \&
  {Lane}}]{2012PASP..124..268M}
{Makarov}, V.~V., {Veillette}, D.~R., {Hennessy}, G.~S., \& {Lane}, B.~F. 2012,
  \pasp, 124, 268, \dodoi{10.1086/664930}

\bibitem[{{Mohayaee} {et~al.}(2020){Mohayaee}, {Rameez}, \&
  {Sarkar}}]{2020arXiv200310420M}
{Mohayaee}, R., {Rameez}, M., \& {Sarkar}, S. 2020, arXiv e-prints,
  arXiv:2003.10420.
\newblock \doarXiv{2003.10420}

\bibitem[{{Planck Collaboration} {et~al.}(2020){Planck Collaboration},
  {Aghanim}, {Akrami}, {Ashdown}, {Aumont}, {Baccigalupi}, {Ballardini},
  {Banday}, {Barreiro}, {Bartolo}, {Basak}, {Battye}, {Benabed}, {Bernard},
  {Bersanelli}, {Bielewicz}, {Bock}, {Bond}, {Borrill}, {Bouchet}, {Boulanger},
  {Bucher}, {Burigana}, {Butler}, {Calabrese}, {Cardoso}, {Carron},
  {Challinor}, {Chiang}, {Chluba}, {Colombo}, {Combet}, {Contreras}, {Crill},
  {Cuttaia}, {de Bernardis}, {de Zotti}, {Delabrouille}, {Delouis}, {Di
  Valentino}, {Diego}, {Dor{\'e}}, {Douspis}, {Ducout}, {Dupac}, {Dusini},
  {Efstathiou}, {Elsner}, {En{\ss}lin}, {Eriksen}, {Fantaye}, {Farhang},
  {Fergusson}, {Fernandez-Cobos}, {Finelli}, {Forastieri}, {Frailis},
  {Fraisse}, {Franceschi}, {Frolov}, {Galeotta}, {Galli}, {Ganga},
  {G{\'e}nova-Santos}, {Gerbino}, {Ghosh}, {Gonz{\'a}lez-Nuevo}, {G{\'o}rski},
  {Gratton}, {Gruppuso}, {Gudmundsson}, {Hamann}, {Handley}, {Hansen},
  {Herranz}, {Hildebrandt}, {Hivon}, {Huang}, {Jaffe}, {Jones}, {Karakci},
  {Keih{\"a}nen}, {Keskitalo}, {Kiiveri}, {Kim}, {Kisner}, {Knox},
  {Krachmalnicoff}, {Kunz}, {Kurki-Suonio}, {Lagache}, {Lamarre}, {Lasenby},
  {Lattanzi}, {Lawrence}, {Le Jeune}, {Lemos}, {Lesgourgues}, {Levrier},
  {Lewis}, {Liguori}, {Lilje}, {Lilley}, {Lindholm}, {L{\'o}pez-Caniego},
  {Lubin}, {Ma}, {Mac{\'\i}as-P{\'e}rez}, {Maggio}, {Maino}, {Mandolesi},
  {Mangilli}, {Marcos-Caballero}, {Maris}, {Martin}, {Martinelli},
  {Mart{\'\i}nez-Gonz{\'a}lez}, {Matarrese}, {Mauri}, {McEwen}, {Meinhold},
  {Melchiorri}, {Mennella}, {Migliaccio}, {Millea}, {Mitra},
  {Miville-Desch{\^e}nes}, {Molinari}, {Montier}, {Morgante}, {Moss}, {Natoli},
  {N{\o}rgaard-Nielsen}, {Pagano}, {Paoletti}, {Partridge}, {Patanchon},
  {Peiris}, {Perrotta}, {Pettorino}, {Piacentini}, {Polastri}, {Polenta},
  {Puget}, {Rachen}, {Reinecke}, {Remazeilles}, {Renzi}, {Rocha}, {Rosset},
  {Roudier}, {Rubi{\~n}o-Mart{\'\i}n}, {Ruiz-Granados}, {Salvati}, {Sandri},
  {Savelainen}, {Scott}, {Shellard}, {Sirignano}, {Sirri}, {Spencer},
  {Sunyaev}, {Suur-Uski}, {Tauber}, {Tavagnacco}, {Tenti}, {Toffolatti},
  {Tomasi}, {Trombetti}, {Valenziano}, {Valiviita}, {Van Tent}, {Vibert},
  {Vielva}, {Villa}, {Vittorio}, {Wandelt}, {Wehus}, {White}, {White},
  {Zacchei}, \& {Zonca}}]{2020A&A...641A...6P}
{Planck Collaboration}, {Aghanim}, N., {Akrami}, Y., {et~al.} 2020, \aap, 641,
  A6, \dodoi{10.1051/0004-6361/201833910}

\bibitem[{{Secrest} {et~al.}(2015){Secrest}, {Dudik}, {Dorland}, {Zacharias},
  {Makarov}, {Fey}, {Frouard}, \& {Finch}}]{2015ApJS..221...12S}
{Secrest}, N.~J., {Dudik}, R.~P., {Dorland}, B.~N., {et~al.} 2015, \apjs, 221,
  12, \dodoi{10.1088/0067-0049/221/1/12}

\bibitem[{{Secrest} {et~al.}(2021){Secrest}, {von Hausegger}, {Rameez},
  {Mohayaee}, {Sarkar}, \& {Colin}}]{2021ApJ...908L..51S}
{Secrest}, N.~J., {von Hausegger}, S., {Rameez}, M., {et~al.} 2021, \apjl, 908,
  L51, \dodoi{10.3847/2041-8213/abdd40}

\bibitem[{{Shen} {et~al.}(2021){Shen}, {Chen}, {Hwang}, {Liu}, {Zakamska},
  {Oguri}, {Li}, {Lazio}, \& {Breiding}}]{2021NatAs...5..569S}
{Shen}, Y., {Chen}, Y.-C., {Hwang}, H.-C., {et~al.} 2021, Nature Astronomy, 5,
  569, \dodoi{10.1038/s41550-021-01323-1}

\bibitem[{{Wojtak} {et~al.}(2011){Wojtak}, {Hansen}, \&
  {Hjorth}}]{2011Natur.477..567W}
{Wojtak}, R., {Hansen}, S.~H., \& {Hjorth}, J. 2011, \nat, 477, 567,
  \dodoi{10.1038/nature10445}

\end{thebibliography}
\bibliographystyle{aasjournal}

\end{document}